# Size-Dependent Structural and Magnetic Properties of Disordered Co$_2$FeAl Heusler Alloy Nanoparticles

**Aquil Ahmad, Srimanta Mitra, S. K. Srivastava and A. K. Das**[†]

Department of Physics, Indian Institute of Technology Kharagpur, India-721302.

[†] Corresponding author: amal@phy.iitkgp.ac.in

## Abstract

Co$_2$FeAl (CFA) nanoparticles (NPs) of different sizes were synthesized by chemical route. The effect of the size of NPs upon the structure and magnetization compared to its bulk counterpart was investigated. The structure and composition were determined from X-ray diffraction (XRD) and electron microscopy. XRD analysis shows that the samples are having single (A2-type) disordered phase. Magnetization measurements suggest that the samples are soft ferromagnetic in nature with very low coercivity. Enhanced magnetic properties like saturation magnetization, coercive force, retentivity, and Curie-temperature are observed with a decrease in particle size. The effect of particle size on hysteresis losses is also discussed. The smallest particles of size 16 nm exhibited the highest saturation magnetization and transition temperature of 180.73 emu/g and 1261 K, respectively. The origin of enhancement in magnetization of Co$_2$FeAl nano-alloy is attributed to the strong Co-Co exchange interaction due to disorder present in the systems.

## 1. Introduction

Heusler Alloys (HAs) referred to ternary intermetallic compounds having general formula X$_2$YZ are generally called full-Heusler alloys. HAs crystallize in the cubic symmetry under space group Fm$\bar{3}$m and fully ordered in ideal L21 phase [1-3]. Full-Heusler alloys may also crystallize in disordered phases like B2 (partially disordered), A2 (fully disordered) and DO3 type phases [4-6]. These alloys show multifunctional properties, like half-metallic ferromagnetism (HMF), shape-memory effect, magneto-caloric effect, and thermo-electric behavior [7-10]. Band-structure calculations divulge that these compounds are half-metallic ferromagnets (HMF), i.e., 100% spin-polarized near Fermi level. Over the past few years, Heusler compounds are considered as the most promising class of materials for spintronics and magneto-electronics applications due to the theoretical prediction of their half-metallicity, high Curie-Weiss temperature (T$_c$) and structural similarity with the binary semiconductors. However, in some alloys, less spin polarization is observed [11,12] due to the structural disorder [13,14] and surface effect [15]. Heusler nanoparticles (NPs) exhibit all the bulk features with different magnitudes and hence it is worth to study the crucial factors, like how size and shape of the nanoparticles affect the structural and magnetic properties of Heusler nanoparticles.

Among all known Heusler alloys, Co$_2$ based Heusler alloys, particularly; Co$_2$FeAl (CFA) attracts more attention due to its unique properties, like high saturation magnetization (M$_s$) and high Curie-Weiss temperature (T$_c$). The experimental lattice parameter of bulk CFA was found to be 5.73 Å and saturation magnetization was obtained around 5.20 $\mu_B$/f.u. at room temperature (RT) [16]. Theoretically, the calculated spin magnetic moment was found to be 4.89 $\mu_B$/f.u. using experimental lattice constant, this theoretical value was slightly less than the predicted value of CFA by the Slater-Pauling (SP) rule, which was 5.0 $\mu_B$/f.u. [10]. The $T_c$ of bulk CFA was predicted ≥1100 K [17]. In recent years, CFA has been investigated extensively, but studies are limited to bulk and thin-films only due to very complicated process of nano-particle synthesis. Phase separation can easily occur in the case of ternary intermetallic alloys due to immiscibility and lattice mismatch; so no particular methodology of synthesis process has been developed till date to achieve nominal phase and stoichiometry. Heusler NPs are also the material of interest as their structural and magnetic properties can easily be tuned by controlling the shape and size of the particles; hence, they are having great importance in the fabrication of nano-devices in spintronics, drug delivery, catalysis and biomedical imaging applications like other magnetic nanostructures mostly based on binary alloys [18-23]. There are very few reports on CFA nanoparticles (NPs); however, the size-dependent studies on $Co_2FeAl$ NPs are not done yet.

Nanostructured CFA was prepared by a hydrothermal method using Teflon-autoclave by Fujun Yang et al. [24]. They have shown the dependency of saturation magnetization ($M_s$) and coercivity ($H_c$) on the reaction time and obtained a maximum saturation magnetization of 189.2 emu/g. Keshab R et al. [25] also prepared CFA nanowires of diameter ranging from 50 to 500 nm with a lattice constant of 5.639 Å. They found that the lattice constant of prepared nanowires was within 1% of the lattice parameter of the bulk system. Hollow CFA NPs were also prepared by using capping agent, namely polyethylene glycol polymer and then annealed at 700 °C with a heating range from 5 to 15 °C per min [26]. They achieved a maximum saturation magnetization of 95 emu/g and $H_c$ of 730 Oe. Al Kanani et al. found Tc above 1000 K for bulk CFA alloy and the saturation magnetization of 5.1 $\mu_B$/f.u. [27]. A. Kumar also chemically prepared CFA nanoparticles in size range of 10-50 nm and reported that nanoparticles were having disordered mixed phases of B2 and A2 types [28]. J. H. Du et al. prepared CFA nanoparticles using co-precipitation method followed by $H_2$ annealing and obtained particles in the size range of 50-400 nm with $M_S$ = 139 emu/g at 300 K [29]. Lana T. Huynh et al. prepared off-stoichiometric CFA nanocrystals by thermal decomposition of the corresponding metal acetylacetonate complexes in the presence of capping ligands, and then annealed in presence of $H_2$ gas [30]. They showed that the saturation magnetization was increased with annealing temperature from 4.0 $\mu_B$/f.u. to 5.3 $\mu_B$/f.u.

In this paper, we report the size effect on structural and magnetic properties of CFA NPs having a mean size of 16 ± 10 nm and 190 ± 9 nm, respectively.

## 2. Experiments

CFA nanoparticles of average sizes of 190 nm and 16 nm within the variation of ± 10 nm were synthesized by co-precipitation and thermal deoxidization method reported by J.H. Du et al. [29] with slight modification. We used the precursor salts: $CoCl_2.6H_2O$ (99%), $Fe(NO_3)_3.9H_2O$ (99%) and $Al_2(NO_3).18H_2O$ (98%). The chemicals from Sigma-Aldrich (American Chemicals Company) were used without any further purification. In a typical preparation of the sample, 0.6 mmol of $Al_2(NO_3).18H_2O$, 1.2 mmol of $Fe(NO_3)_3.9H_2O$ and 2.4 mmol of $CoCl_2.6H_2O$ were dissolved in 50 ml methanol and then dried for more than 12 hours in 100 °C. Obtained dried powder was further annealed at 800 °C for 3 hours and 12 hours in presence of H₂ atmosphere to deoxidize the powder.

Phase identification was studied by high-resolution x-ray diffractometer (HR-XRD) using Cu Kα radiation from (i) Bruker D2 Phaser x-ray diffractometer and (ii) PANalytical's x-ray diffractometer. The morphologies of prepared NPs were analyzed by MERLIN field-emission gun scanning electron microscope (FESEM), which was operated at 200 kV. The colloidal nanoparticles were pre-sonicated in acetone, a small drop was placed in a small Si substrate and then gold coating was used to prepare the sample for FESEM, this drop was further placed on a carbon supported Cu (Copper) grid for the transmission electron microscope (TEM). To determine the purity and composition energy-dispersive x-ray analysis (EDAX) was used. DC magnetization measurements were carried out using a physical property measurement system (PPMS) of Cryogenic Limited, UK and high-temperature vibrating sample magnetometer (VSM) of LAKESHORE, USA.

## 3. Results and discussions

### 3.1 X-ray diffraction (XRD)

XRD patterns of all the prepared Co$_2$FeAl (CFA) samples annealed at 800 °C for (a) 3 hours and (b) 12 hours are shown in Fig. 1. The pattern consists of three clear characteristic peaks at 44.96°, 65.50°, and 82.92°. These peaks are indexed as (220), (400) and (422) reflections of pure Co$_2$FeAl phase according to the Joint Committee on Powder Diffraction Standards (JCPDS) data. It is observed that the peaks are relatively displaced towards higher 2θ values compared to their bulk counterpart, so the NPs are pure in phase. Furthermore, XRD patterns reveal the crystalline nature of the samples. The calculated lattice constants are of 5.695 Å and 5.718 Å for 3 hours and 12 hours annealed samples, respectively, which are comparable to the bulk lattice constant [16]. From the literature, it is well known that the ideal $L2_1$ structure is recognized if (111) and (200) superlattice peaks are present in XRD patterns [31], which are not present in our XRD spectra as shown in Fig. 1. These superlattice peaks are much weaker for the elements belong to a same period in the periodic table and hence intensities of these peaks could not be easily detectable if the transition elements in the Heusler-compounds belong to the same period. These peaks are also not existed if samples have A2-type disordered structure. Subsequently, the Al atom in the periodic table belongs to a different period than that of Co or Fe, so our samples exhibited an A2 type disordered phase [7].

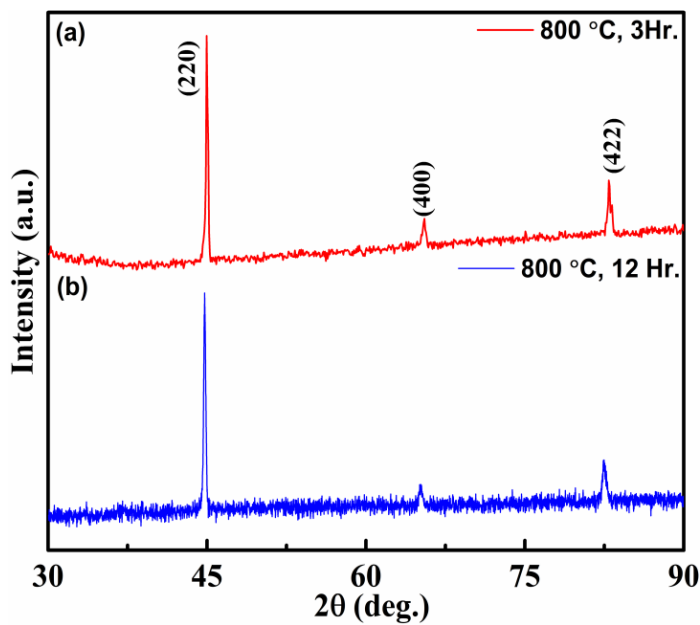

**Fig. 1:** XRD patterns of the Co$_2$FeAl (CFA) nanoparticles annealed at 800° C for (a) 3 hours and (b) 12 hours.

### 3.2 Field emission scanning electron microscopy (FESEM) and transmission electron microscopy (TEM)

Field emission scanning electron microscopy (FESEM) images in Fig. 2 are indicating that all the particles are spherical in shape and agglomerated. The agglomeration may be due to the magnetic nature of the particles.

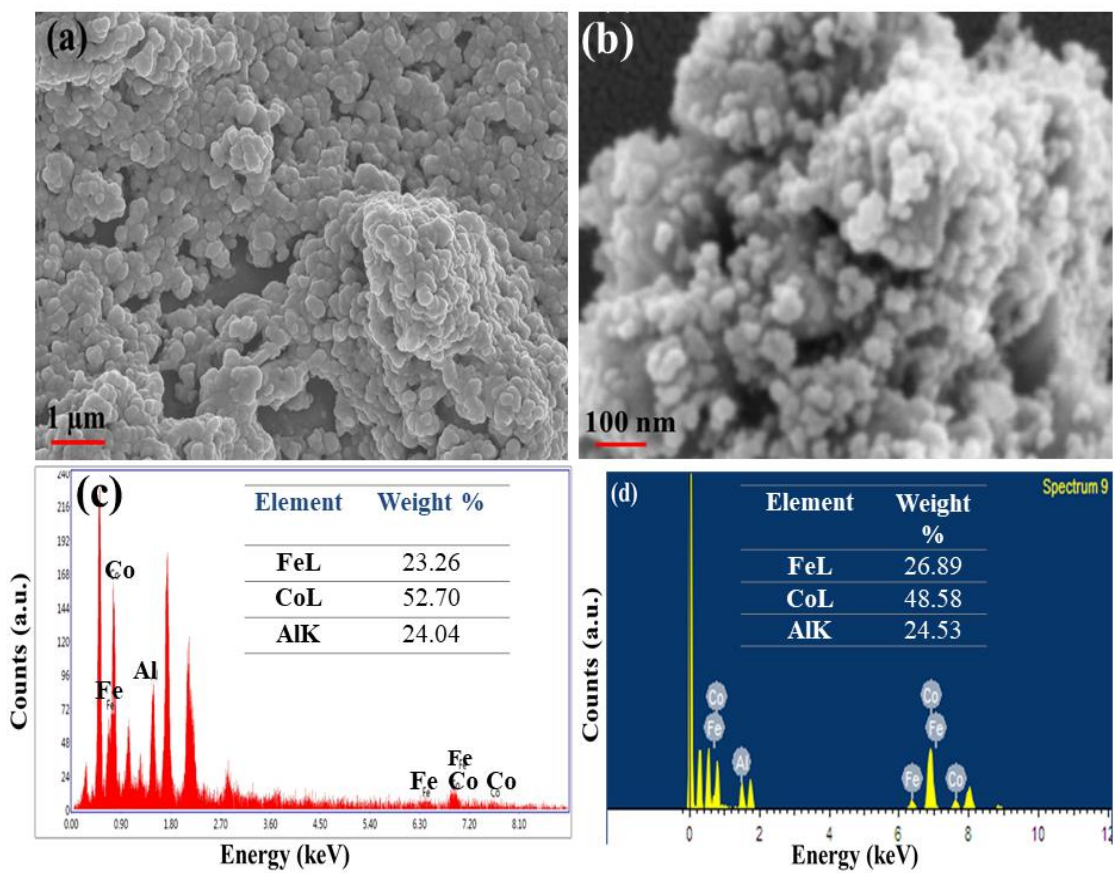

**Fig. 2:** FESEM images of the $Co_2FeAl$ (CFA) nanoparticles annealed at 800° C for (a) 3 hours and (b) 12 hours. EDAX spectrum of CFA annealed at 800 °C for (c) 3 hours and (d) 12 hours.

To confirm the purity and composition we have performed energy-dispersive x-ray analysis (EDAX) shown in Fig. 2 (c) and (d), which reveal that there are no foreign elements. Note that the extra peaks in the spectrum are from Si substrate and Au as the gold coating was done on the Si substrate during the sample preparation. This composition examination demonstrated that the deviation from the stoichiometric composition of all the samples is insignificant.

To analyze particles further we have performed transmission electron microscopy (TEM) imaging along with selected area electron diffraction (SAED) and shown in Fig. 3. The SAED patterns encompass concentric rings with some dots, which is indicating the single crystalline nature of the nanoparticles but randomly oriented in powder form. Therefore, the NPs are essentially crystalline in nature. Primary four rings of the SAED patterns are indexed with (hkl) as shown in Fig. 3 and those (hkl) values are also in good agreement with the XRD results. We analyzed all particles by using Image J software and the histogram of size distribution of nanoparticles is shown in Fig. 3 and fitted with a Gaussian profile, which gives the average particle sizes of 190 ± 9 nm and 16 ± 10 nm for 3 hours and 12 hours annealed samples, respectively.

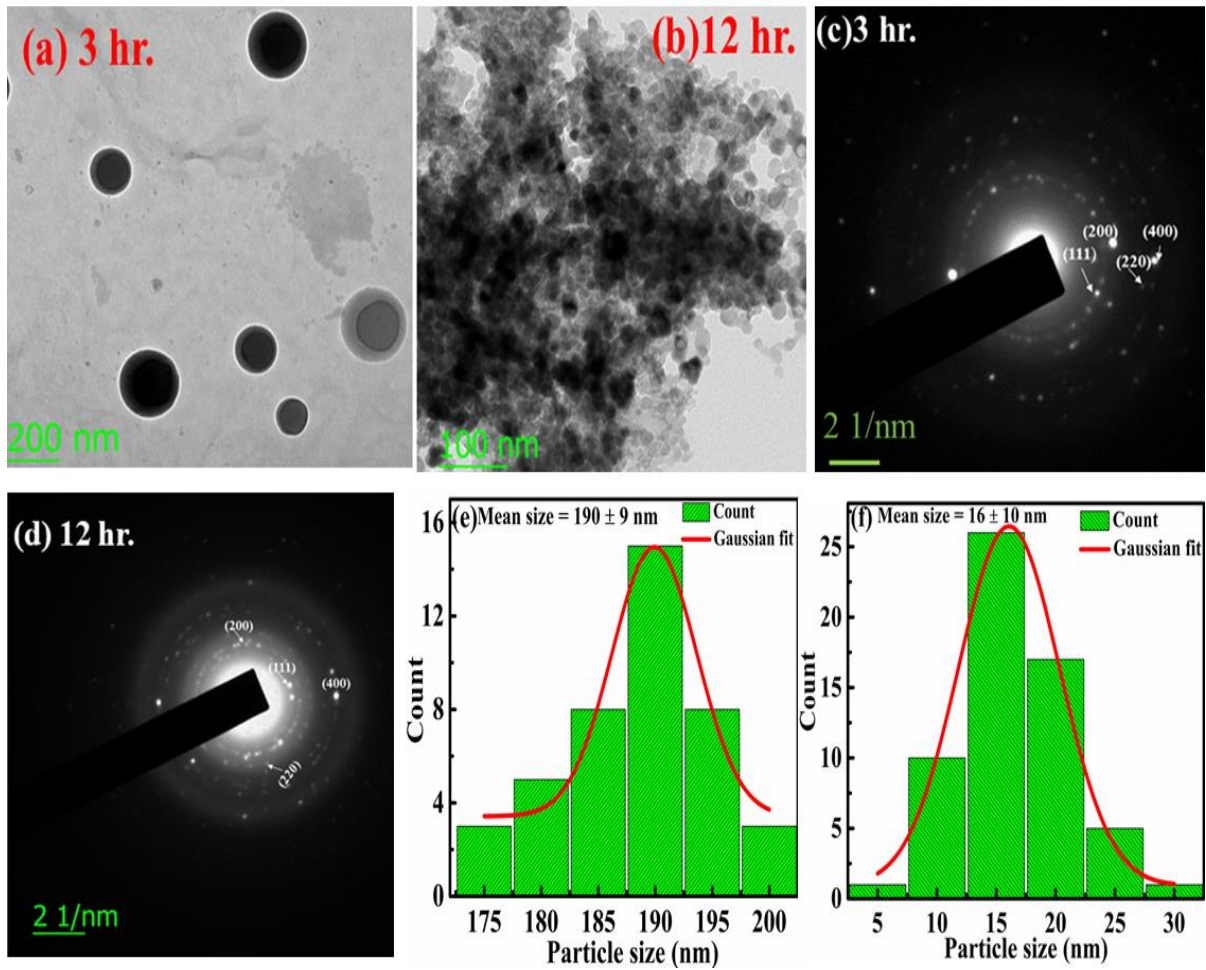

**Fig. 3:** TEM images of Co$_2$FeAl (CFA) NPs annealed at 800 °C for (a) 3 hours and (b) 12 hours. SAED patterns of CFA annealed at 800 °C for (c) 3 hours and (d) 12 hours, with an index of initial four rings. (e) and (f) show the histogram of particle size distribution for 3 hours and 12 hours annealed samples, respectively.

### 3.3 Effects of particle size on magnetic properties

Magnetization as a function of the magnetic field was measured at low temperature (5 K) and at room temperature (RT). It is seen that the particles are ferromagnetic in nature and the saturation magnetization (M$_s$) is increased with the decreasing of particle size. The M$_s$ are 147.62 emu/g and 180.73 emu/g for 190 nm and 16 nm sized NPs, respectively. The same tendency is observed in case of coercivity (H$_c$) and remanence (M$_r$); both coercivity and remanence are also increased with decrease in particle size at 5 K. The H$_c$ are 121.13 Oe and 175.50 Oe and the M$_r$ are 2.63 emu/g and 8.05 emu/g for 190 nm and 16 nm-sized nanoparticles, respectively at 5 K. Similar trend are also observed at room temperature. Moreover, the saturation magnetization

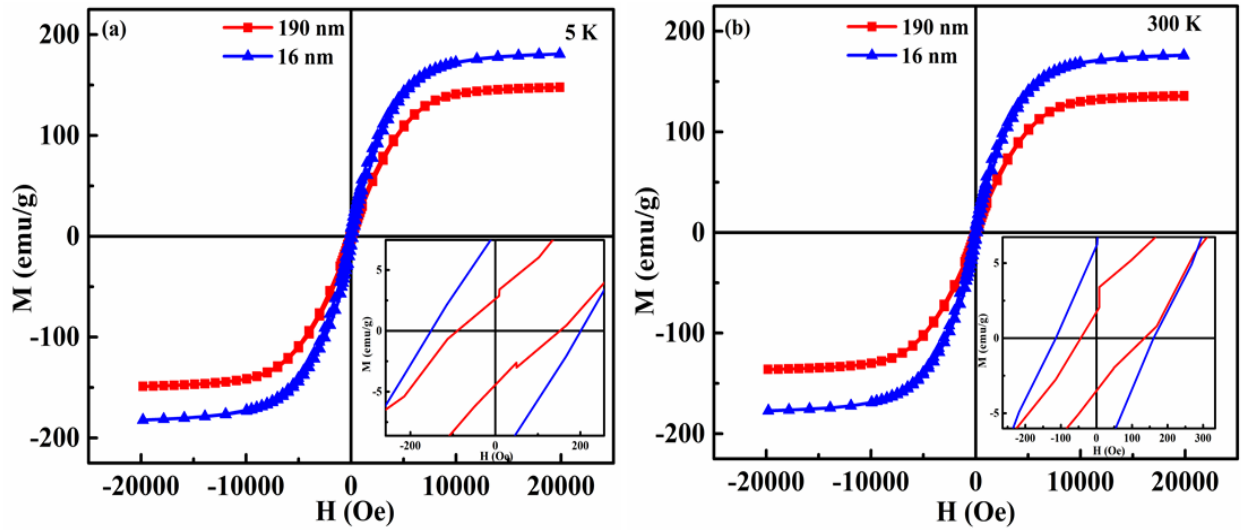

**Fig. 4:** Field dependence of magnetization at (a) 5 K and (b) 300 K. Insets show the magnified curves around zero magnetic field.

(M$_s$) is much stable even up to RT, which has great importance for device application. From Fig. 4, it can be seen that all particles are soft ferromagnets due to very small coercivity (H$_c$). The particles having the smallest size of 16 nm exhibited a highest magnetic moment of 180.73 emu/g (~6.5 µ$_B$/f.u.), which is also comparable with the M$_s$ value of 189.2 emu/g reported by F. Yang et al. [24]. This value is much larger than the value reported for the bulk system (see in table 1). The enhancement in the magnetic moment and also in Curie temperature compared to bulk is attributed to the disorder (A2-type) present in our system, which leads to strong magnetic exchange interaction and as a result large magnetic moment and Curie temperature. The strong exchange interaction is basically due to the Co-Co exchange interactions. As in fully-disordered CFA lattice, the Co atom will have Fe or Al as its nearest neighbor which can lead Co-Al-Co exchange interaction [7]. According to Slater-Pauling (SP) rule, the theoretical bulk saturation-magnetization value for CFA is 5.0 µ$_B$/f.u. [10]. In our case, for 190 nm-sized particles saturation magnetization value is in good agreement with the theoretical bulk value. When we go to nano-regime of particle size of 16 nm, this value becomes 180.73 emu/g, at 5 K and maintained even up to room temperature. A comparison of all values obtained at 5 K and 300 K has shown in Table 1. We have also measured magnetic hysteresis loops at 100 K and 200 K for 190 nm and 16 nm-sized particles and the obtained values are shown in Fig. 6.

**Table. 1:** Saturation magnetization, remanence, coercivity, Curie temperature ($T_c$) of Co$_2$FeAl nanoparticles of sizes 190 nm and 16 nm.

| Particle size (nm) | Sat. mag. M$_S$ (emu/g) (5K) | Remanence M$_r$ (emu/g) (5K) | Coercivity H$_c$ (Oe) (5K) | M$_S$ (emu/g) (300 K) | M$_r$ (emu/g) (300 K) | H$_c$ (Oe) (300 K) | Curie temperature T$_c$ (K) |
|---|---|---|---|---|---|---|---|
| 190 | 147.62 | 2.63 | 121.13 | 135.80 | 1.75 | 89.0 | 1220 |
| 16 | 180.73 | 8.05 | 175.50 | 176.15 | 6.12 | 137.4 | 1261 |
| **Bulk** | - | - | M$_s$ 5.2 µ$_B$/f. u. [16] | | - | - | ≥1100K (predicted), 1170 K (exp.) [17], [32] |
| **Slater-Pauling (SP) M$_s$ value** | | | 5.0 µ$_B$/f.u [10] | | | | |

Fig. 5 shows the variation of magnetization with respect to temperature under an applied field of 100 Oe with standard ZFC (zero-field cooling) technique. The inset shows dM/dT versus T plots for particles sized of 190 nm and 16 nm. From the temperature dependence of the magnetization curve, it is clear that all samples undergo ferromagnetic (FM) to paramagnetic (PM) phase transition at Tc equals to 1220 and 1261 K, respectively. Thus, T$_c$ increases with the decrement of particle sizes, which is very pronounced. We are not aware of any measurement of Curie-temperature (T$_c$) for CFA nanoalloy, so we could compare our experimental values with its bulk counterpart and it is much larger than its bulk value measured experimentally and predicted theoretically (see in Table.1). However, Changhai Wang et al. studied Co-Ni-Ga Heusler nanoparticles and reported highest T$_c$ of 1174 K for 84 nm sized particles [33].

The largest bulk magnetic moment and Curie temperature were predicted for Co$_2$FeSi according to SP rule and Fecher et al. found an experimental magnetic moment and T$_c$ of 5.97 µ$_B$/f.u. and 1100 K, respectively for Co$_2$FeSi alloy [34]. However, we found highest M$_s$ for Co$_2$FeAl nanoparticles, which is higher than 6 µ$_B$/f.u. Our sample with mean size of 16 nm, exhibited highest T$_c$ among all Heusler alloy till date. From these results, we can conclude that $Co_{2+x}Fe_{1-x}Z$ (where Z is Al and Ga) in the A2 phase would be the best among all Heusler alloys. It is also interesting to investigate the size effect on hysteresis losses, which is an important parameter for evaluating the performance of magnetic refrigerant materials, where minimization of hysteresis losses is better for higher refrigerant capacity (RC). Therefore, the magnetic hysteresis of ferromagnetic materials described by magnetic coercivity is worth to study.

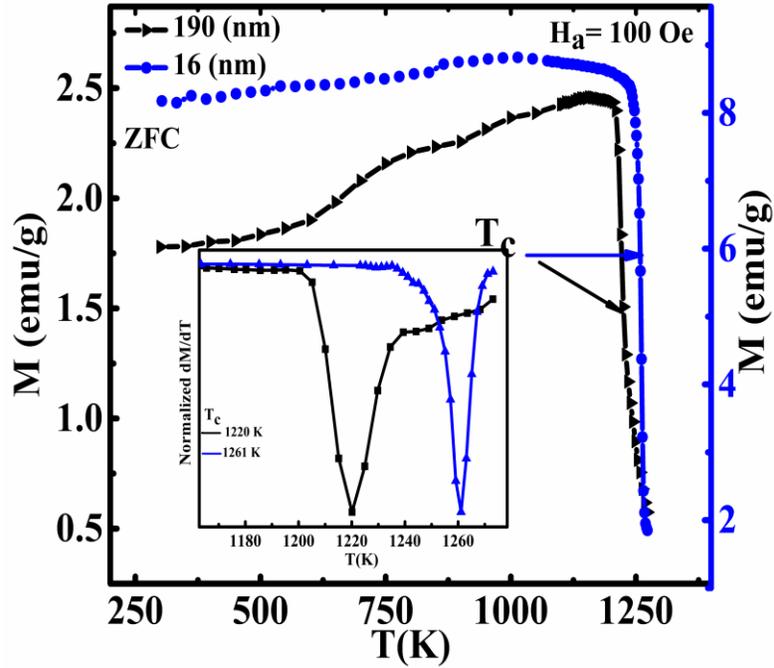

**Fig. 5:** Temperature dependence of magnetization, Inset shows dM/dT near the transition region as a function of temperature for the particles of sizes of 190 nm and 16 nm.

From Fig. 6, It has been observed that coercivity ($H_c$) and remanence ($M_r$) is increased as nanoparticle size goes from 190 nm to 16 nm. Moreover, it is evident that at any given - temperature, magnetization, and size of the nanoparticles are inversely related and hence when the particle size is reduced, the magnetization is increased. The coercivity and remanence of the ferromagnetic NPs are strongly dependent upon the temperature. This happens because when temperature increases, more thermal energy is provided to the system. Hence, all individual electron spins will be in higher energy state, so pointing randomly with respect to neighboring spin and hence magnetization will be reduced; as a result, a small field is required to reduce the remanent magnetization to zero, therefore leading to reduction in coercivity [35,36].

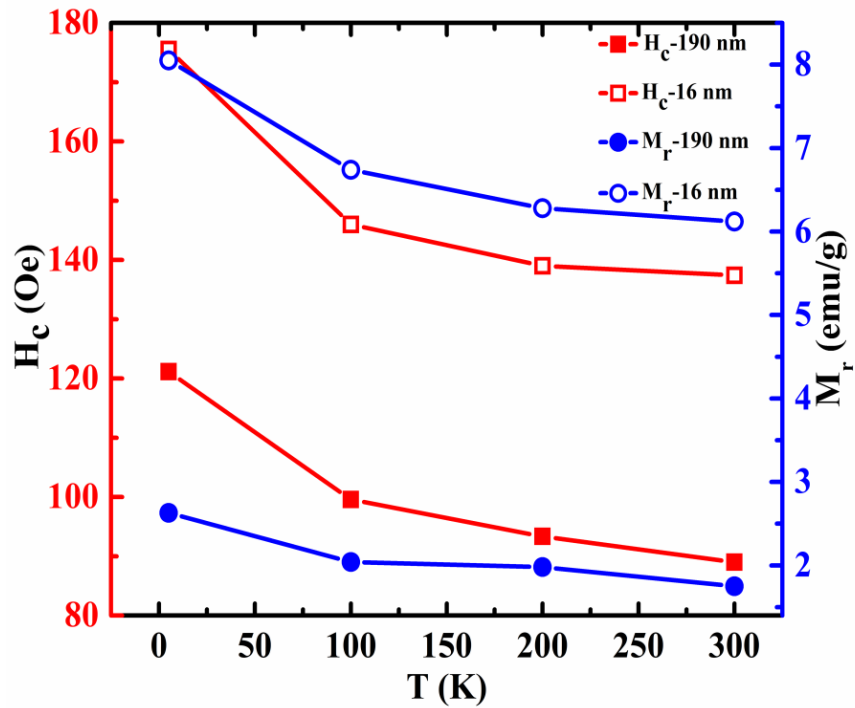

**Fig. 6:** (color online) Coercivity (Hc) and remanence (Mr) plot of CFA-NPs as a function of temperature for the particles sizes of 190 nm and 16 nm.

Here, we have shown that the magnetic moment and Curie temperature can be tuned significantly by controlling the shape and size of the nanoparticles.

## 4. Summary

In summary, we successfully synthesized different size of nearly monodispersed $Co_2FeAl$ nanoparticles and obtained the smallest mean size of 16 nm. All samples are pure in phase and have crystallized in the A2 type of disordered phase. Our studies clearly reveal that the disorders, which are present in our CFA nanoalloys, lead to enhance saturation magnetization and Curie temperature ($T_c$) when particle sizes are reduced. We found the maximum saturation magnetization of 180.73 emu/g (~6.5 $\mu_B$/f.u.), and Curie temperature of 1261 K for the particles with a mean size of 16 nm. We found the drastic change in magnetic moment and Curie temperature ($T_c$) with the size of the nanoparticles as compared to its bulk counterpart. The study of hysteresis losses also suggests the vitality of these low-cost materials for magnetic refrigeration technology. Our finding of CFA alloy having a high magnetic moment and Curie temperature would be interesting for fundamental science and important for device applications.

## 5. Acknowledgments

Aquil Ahmad sincerely acknowledges the University Grant Commission (UGC) Delhi, India for providing fellowship for Ph.D. work. A. K. Das acknowledges the financial support of DST, India (project no. EMR/2014/001026). We also acknowledge the Central Research Facility of IIT Kharagpur and the use of PPMS facility in the Department of Physics of IIT Kharagpur. We are also thankful to Dr. Koushik Biswas, Department of Metallurgical and Materials Engineering, IIT Kharagpur for providing $H_2$ annealing facility.